\documentclass[preprint2]{aastex}
\newcommand{\NaHI}{\ion{Na}{1}/\ion{H}{1}\ }
\newcommand{\kms}{\ensuremath{\mathrm{km}\ \mathrm{s}^{-1}}}

\begin{document}

\title{The DDO IVC Distance Project:  Survey Description and the Distance
       to $\mathrm{G}139.6+47.6$}

\author{Christopher R. Burns, Christopher Tycner, Megan McClure, Kris Blindert, 
        Rosemary McNaughton, Michael D. Gladders, and
        Allen Attard}
\affil{Department of Astronomy and Astrophysics and the David Dunlap 
       Observatory, 60 St. George Street, University of Toronto, Toronto ON,
       M5S 3H8, Canada}

\begin{abstract}
We present a detailed analysis of the distance determination
for one Intermediate Velocity Cloud (IVC $\mathrm{G}139.6+47.6$) from the 
ongoing DDO
IVC Distance Project.  Stars along the line of sight to $\mathrm{G}139.6+47.6$
are examined for the presence of sodium absorption attributable to the
cloud, and the distance bracket is established by astrometric and 
spectroscopic 
parallax 
measurements of demonstrated foreground and background stars.
We detail our strategy regarding target 
selection, observational
setup, and analysis of the data, including a discussion of wavelength
calibration and sky subtraction uncertainties.  We find a distance estimate 
of $129 \pm 10$ pc for the lower limit and 
$257^{+211}_{-33}$ for the upper
limit.  Given the high number of stars showing absorption due to this
IVC, we also discuss
the small-scale covering factor of the cloud and the likely significance of 
non-detections for
subsequent observations of this and other similar IVC's.  Distance measurements
of the remaining targets in the DDO IVC project will be detailed in a companion
paper.
\end{abstract}


\section{INTRODUCTION}

As with many objects in the Universe, 
Intermediate Velocity Clouds (IVC's) are poorly understood due to a lack of 
distance estimates.  
Many of the interesting physical properties of the IVC's (mass, density, volume, etc.)
depend on the distance, and
without
these properties it is difficult to constrain models concerning their 
origins, evolution, or relative importance to the interstellar medium
\citep{Wakker:1997}.

There are several ways to determine the distances to IVC's, which have been
used to varying degrees of success 
\citep[see][and references therein]{Gladders:1998}.  
One of these methods stands out as
particularly successful in determining both upper and lower limits
on the distances to the clouds:  the absorption line method.  In this method,
stars along
the line of sight to the cloud are
used as distance markers.  If a
star is behind the cloud, one hopes that metallic lines at the cloud's
systemic velocity can be detected in the star's spectrum.  By determining 
whether stars are behind
the cloud, we set upper and lower limits on its distance.  
Astrometric or spectroscopic
parallax can then be used to determine the distances to the observed stars to
bracket the cloud's distance.
Such a technique, unfortunately,  requires a large investment of
telescope time \citep{Wakker:1997}. 

We began the DDO IVC Distance Project in the summer of 1997 in the hopes
that a reliable distance to the Draco cloud could be determined
\citep[see][]{Gladders:1998}.  The success of this project soon led to
distance determinations for other IVCs \citep[see][]{Clarke:1999}.
In this paper, we use the cloud $\mathrm{G}139.6+47.6$ 
as a case
study in order to present our methodology as a precursor to the publication 
of the entire data set, forthcoming in a companion paper.  
This cloud's Local Standard of Rest velocity is $v_{\rm LSR} = -12\ \kms$ and
is one of the lowest in our sample, which range from $3.5 \ \kms$ to
$-47\ \kms$.  Traditionally, IVC's are distinguished by 
$\left|v_{\rm LSR}\right| < 90\ \kms$ and $\mathrm{G}139.6+47.6$, having a
relatively low velocity, might be more appropriately labeled a Low Velocity
Cloud (LVC).  However, as we are not discussing the origins or kinematics of
these clouds, we will not make this distinction here.

\section{OBSERVATIONS}

\subsection{Target Selection}

We have taken our target list of clouds from the sample of 
\citet{Heiles:1988}.  Among the 26 clouds, we selected 16 to observe
 that are
at high galactic latitudes and are observable from the David Dunlap
Observatory.
Several of these clouds have been grouped together in pairs, as 
they do not seem to be separated either in angular or velocity space.
We are thus left with a sample of 11 cloud complexes.

With the clouds selected, we then require good stellar
targets along their lines of sight.  We consider a ``good'' target
to be a star with $B - V < 0.35 $, as we want early-type stars 
with as few stellar features as possible so
that absorption features can be easily identified as originating in the
cloud rather than the stellar photosphere.
We use the Tycho Catalogue
\citep{Hog:1997} and the USNO Catalog \citep{Urban:1998} to select our 
targets.  Cloud sizes are estimated 
through both the
\ion{H}{1} atlas  of \citet{Hartmann:1997} and through IRAS 100 
$\mu$ maps (see Figure \ref{fig:targets}).
We then search the Tycho Catalogue, using the online tools, for stars with
positions within the cloud and with the appropriate colors.  When 
the positions of the stars have been determined, we build an \ion{H}{1} 
column density profile in 
velocity space at the position of each star (see Figure \ref{fig:HI+vel}).  
Assuming hydrogen or IRAS emission is a good indication of sodium 
absorption (the validity of this will
be discussed in \S \ref{detect-non-detect}), we can then
determine which stars
represent good probes of the cloud.  Also, using the $B-V$ color index, the
stars' distances can be estimated, which further helps in target
selection.  Figure \ref{fig:targets}
shows the IRAS 100 $\mu$ emission for
the IVC cloud $\mathrm{G}139.6+47.6$ as well as the positions of the 16 
target stars that
satisfied the above criteria.

\subsection{Observational Setup}

Since the beginning of this project, we have 
been assigned a
total of 137 nights, of which 63 have produced useful spectra for at least
a portion of the night.  To date we have
observed 86 stars along lines of sight to eleven different IVC's.

Nominally, each target star is observed using two observational setups:
a sodium setup and a classification setup.  Both use the
David Dunlap Observatory (DDO) 1.88~m telescope and Cassegrain spectrograph.  
The classification setup is used to classify each star so that its distance can
be determined.  The sodium setup is used to observe the \ion{Na}{1} doublet to
determine whether the star is behind the cloud or not.  The sodium setup is at
higher resolution than the classification setup, and so it represents the bulk 
of the
observations.  In addition to our program stars, spectroscopic standards are
observed in both the sodium and classification setups.  We
observe standards in the sodium setup in order to generate template spectra
 for use with the cross-correlation techniques, which are detailed in \S
\ref{DetectNonDetect}.  

\subsubsection{Sodium Setup}

The sodium setup yields a wavelength coverage of approximately 
5800--6000 \AA\ with a resolution of $0.43$ \AA.
With this setup, we cannot resolve any sodium absorption  caused by the cloud 
(our velocity resolution is $22\ \kms$ at the \ion{Na}{1} doublet); however, this 
is not necessary for our purposes (see \S \ref{DetectNonDetect}).  We 
observe targets down to a 
magnitude of 12.5, for which it takes 4 hours to achieve a signal to noise 
ratio 
of 50.
The wavelength region of interest is significantly affected 
by water vapor absorption lines, so the observation of telluric standards
is important.  Typically, we observe three to four telluric standards 
throughout the night.  These standards are later used to
subtract the telluric lines from the sodium spectra.

\subsubsection{Classification Setup}
For the classification setup, we cover the range 3800--4400~\AA\ at a
resolution of $2.5$ \AA.  
We obtain classification spectra of any stars which show definite 
absorption or lack of absorption by the cloud and that do not have
Hipparcos distance measurements.
Classification spectra of standards taken from \cite{Garcia:1989} are also 
obtained,   
covering spectral types B0 to G5 and luminosity classes I to V.  These 
standards
are used to determine the distances to several program stars using
spectroscopic parallax.

\section{DATA ANALYSIS}

\label{sect:spectrum}

The bulk of the data reduction is now done through a processing pipeline
developed using standard IRAF tasks.  
This pipeline allows
us to quickly reduce a night's data and also ensures a consistent
procedure for all data.  The
rapid turnaround allows greater precision in subsequent
target selection. Further, with over 
20 observers working on the same project, a centralized 
repository of data is necessary.  Such a database, implemented in
HTML, was devised and became an invaluable resource.

The data are de-biased and flat-fielded using standard methods.  Each 
two-dimensional spectrum is then rectified to a common wavelength scale.
With the DDO's close proximity to one of North America's largest urban centers,
one must be very careful in the subtraction of the sodium sky lines.  Even 
though most
of the light from high-pressure sodium lights is self-absorbed,
the two sodium lines from low pressure lamps are
quite bright (see Figure \ref{fig:skylines}).  Great care is taken to
ensure proper rectification of the 2-D spectra and hence proper sky line
subtraction.

Lastly, we remove the telluric lines using the spectrum of the telluric 
standard that is observed the closest in time to the program star.  This is
done by normalizing the reduced, continuum-subtracted 1-D telluric spectrum 
and then correcting for the difference in airmass.
Figure
\ref{fig:tell} shows a typical spectrum of the telluric standard
$\mathrm{HD}177724$.
Two telluric standard stars were chosen due to their lack of contaminating 
inter-stellar 
features in the wavelength region of interest: $\mathrm{HD}177724$ (A0Vn, 
$m_{\rm V}=2.988$) and 
$\mathrm{HD}120315$ (B3V, $m_{\rm V}=1.852$).
In order to sample the slit effectively without saturating the CCD, 
neutral density filters were used. 

In order to test for the wavelength calibration stability, the sky lines
are extracted from the dispersion-corrected 2-D spectra of every program
star.  Using the same cross-correlation techniques outlined in \S 
\ref{DetectNonDetect}, the
shift in the sky \ion{Na}{1} emission lines is computed.  The average shift
is $0.35~\kms$ with a standard deviation of approximately $1~\kms$, which 
is consistent with no systematic shift and demonstrates the stability of
the wavelength calibration.

\section{DISTANCE DETERMINATION}
There are three steps to the determination of a cloud's distance:  1) 
testing for non-stellar contribution in \ion{Na}{1} absorption lines,
2) assigning each
star to the background or foreground, and 3) computing the distance to 
each star.  Each of these steps is described
in detail below.

\subsection{Detection of Absorption by the Cloud}
\label{DetectNonDetect}
The most important step in the determination of a cloud distance is the
detection (or non-detection) of \ion{Na}{1} from the cloud in the spectrum of
stars along the line of sight.  Ideally, one would like to use early-type
stars to minimize the number of stellar lines in the vicinity of the 
\ion{Na}{1} doublet, and one would like to resolve the absorption due to the 
cloud.
However, to maximize the number of target stars for this project, we have 
chosen 
later-type stars  and lower resolution than other authors \citep{Benjamin:1996}.  
Because moderate dispersion allows for a larger sampling of
the underlying stellar spectrum, we can use cross-correlation techniques to
decouple absorption due to the stellar envelope and any possible absorption
due to the IVC.  The technique is as follows.

For each star in our sample, we estimate the spectral class using the
sodium part of the spectrum, or the classification spectrum if this has
been obtained already.  Based on this classification, we build a template
sodium spectrum by measuring the central wavelength and equivalent width
of all significant lines in the sodium spectrum of a standard star with 
similar spectral class.  The template spectrum is composed of a flat continuum
with absorption lines at the same wavelengths and with the same equivalent
widths (though with FWHM less than the resolution of the real spectra) as
would be measured at rest ($v_{r} = 0$).  Using the IRAF task 
{\tt rvcorrect},
each program star's spectrum
is corrected to remove the velocity of the Earth and of the Sun relative
to the Local Standard of Rest (LSR). Measurement of the Doppler 
shift in the absorption features relative to their rest wavelengths
will therefore yield the LSR velocity of 
the star.

Once the template spectrum has been produced, we cross-correlate it with
the spectrum of the program star in two separate regions.  The first region,
which we shall term the {\it stellar region}, is the whole sodium spectrum but
excluding the region around the \ion{Na}{1} doublet.  The second region we shall
term the {\it sodium region}, which contains only the sodium doublet.  The 
cross-correlation of the stellar region
with the template spectrum gives us an estimate of the radial velocity 
and the average width
of the stellar features.  The cross-correlation of the sodium region with
the template spectrum gives us an estimate of the velocity and average 
width of both
the stellar features and the interstellar features if both are present
in the region.  By comparing the
cross-correlations, one can determine whether or not there are any \ion{Na}{1}
features due to interstellar absorption and, therefore, whether
the star is behind the cloud.  If the stellar features are too narrow and too
close to the velocity of the IVC, the star is rejected.  Figure
\ref{fig:cross_corr} shows the cross-correlations for the 4 
program stars we have identified as background stars.
The left panels show the \ion{Na}{1} region of the spectra and the right
panels show the cross-correlations with the template spectra.  

We consider
a star to have a sodium detection if the cross-correlation of the \ion{Na}{1}
D lines peaks near the velocity of the cloud and either there are no
significant stellar lines or any stellar lines are at a significantly 
different velocity than that of the cloud.  There are cases where the 
velocity of
the stellar lines is too close to the velocity of the cloud, yet the 
cross-correlation peak of the stellar lines is significantly broader 
than that of the sodium lines, which is suggestive of absorption due to
the cloud.  In these cases, we label the stars as
possible detections.  Likewise, there are cases where the sodium 
and stellar lines have identical cross-correlation profiles near the velocity
of the cloud and we label these as possible non-detections (see Table 
\ref{table:program}).

\subsection{Detection vs. Non-detection}
\label{detect-non-detect}
While the detection of absorption features due to the cloud in the spectrum
of a star is strong evidence that the star is behind the cloud, the
reverse is not necessarily true.  A non-detection for a star can either
mean the star is in the foreground, or it can simply mean the star is 
behind the cloud but the column density of \ion{Na}{1} is not sufficient to 
produce detectable features in its spectrum.   Two uncertainties arise 
at this point.  The first is the covering factor of the cloud on small
angular scales.  The \ion{H}{1} column density is sampled from a relatively 
large
beam, whereas our stars sample sodium absorption on 
an extremely small angular scale.  It is therefore possible that small-scale
variations in the covering factor are unresolved by the large \ion{H}{1}
beam.  The second uncertainty is the actual abundance of \ion{Na}{1} relative to
\ion{H}{1}.  In the absence of any \ion{Na}{1} absorption, the ratio of \NaHI is
uncertain and could be very low.  Therefore, clouds with no detections may
very well have low \ion{Na}{1}/\ion{H}{1}. 
This problem could be resolved by extrapolation from other cloud metallicities,
once they are determined to some degree of accuracy.
An example of low \NaHI could be $\mathrm{G}86.5+59.6$,
which has eluded all our attempts at detecting \ion{Na}{1} absorption.  Our current
lower limit to the distance to the cloud is 430 pc \citep{Clarke:1999}.  
However, its rather
large radial velocity of $-39~\kms$ may imply a large distance based on
the terminal drag model of \citet{Benjamin:1996},
which predicts a linear correlation between the velocity perpendicular to the
galactic plane and height above the plane.

Recent measurements of IVC sight-lines have shown that \NaHI
varies on the order of 70 per cent over 30 arcsec and by as much as a factor
of 2 over 1 arcmin \citep{Smoker:2002}.  Given the large angular distances between our target
stars, we cannot expect the \NaHI ratios obtained from our
detections to be constant, nor can we use them to accurately predict the
expected equivalent widths of \ion{Na}{1} in other targets.  This makes the
identification of foreground stars more uncertain.  Nevertheless, when 
evaluating the
significance of a non-detection, the lowest observed 
EW(\ion{Na}{1})/$\rm N_{HI}$ 
for the cloud is used to predict the sodium absorption one would expect if 
the star were behind the cloud.  If this absorption is detectable by our
instruments yet is not observed, we conclude the star is in the foreground.  
Figure 
\ref{fig:fake_Na} shows the expected sodium absorption in the spectrum of
$\mathrm{TYC}\ 4151\ 1452$ if the star were behind the cloud. The 4 
spectra were computed using the 
EW(\ion{Na}{1})/$\rm N_{HI}$ obtained from
our 4 detections.  

To address the issue of covering factor, we examine the distance distribution
of our target stars.
Of the 16 program stars, 4 have definite detections 
of \ion{Na}{1} absorption.  By 
obtaining
distances to all these stars, one can now start to answer the question of
covering factor, at least for this particular cloud complex.
We wish to determine the probability that a background star will show
sodium absorption due to the cloud given the \ion{H}{1} column density along
its line of sight.  This can be determined in the following way.  We
first determine the best upper limit on the distance to the cloud (i.e.,
the closest star with a detection).  We then look for any stars which
could be farther away and count the number of detections versus the
number of non-detections.  If we
find a non-detection which is farther away than a detection, this implies
a covering factor less than unity.

If we leave out the stars for which the detections and non-detections are of
low confidence, then there
are 4 stars with detections behind the nominal position of the cloud and
one non-detection in front.
If we include the possible detections and
non-detections, then there are 7 detections behind the furthest non-detection
and two non-detections in front of the closest detection.
In both cases, this is consistent with a covering 
factor of unity.

\subsection{Stellar Distance Determination}
\label{Sect:StellarDistance}
For the determination of the stellar distances, we use either
trigonometric parallax from Hipparcos or spectroscopic 
parallax using spectroscopic classification obtained either in the
literature or from our own observations.
Clearly, the faintest and 
farthest stars require spectroscopic classifications to obtain accurate
distances.  However, since the goal is to determine the best distance to
any given \ion{H}{1} cloud, we are more concerned with accurately determining the
distance to those stars that are closest to the cloud.  

To determine distances based on the spectral type of the star, we use the 
the spectral class and absolute magnitude calibration of \citet{Corbally:1984}
to assign an absolute magnitude to each classified star. 
The Tycho apparent magnitude is then used to 
compute the
distance modulus.  No correction for interstellar absorption is made
as it is negligible at this galactic latitude and longitude, being near
the Lockman Hole.  We also ignore any
extinction that may be caused by the cloud itself, as this is likely small
\citep{Stark:1997} and can only reduce the distance of the background stars, 
thereby improving
our distance estimates.  In assigning uncertainties to our distance estimates,
two factors contribute the most:  the uncertainty in the classification of
the spectra and the intrinsic scatter in the absolute magnitude versus
spectral type relation.  For the former, we assume an uncertainty of plus
or minus one spectral class.  For the latter, we take $\pm 0.7\ \mathrm{mag}$
\citep{Jaschek:1996}.  For the F-type stars, there is also a larger
uncertainty in establishing the luminosity class, which translates to a very
large distance uncertainty.  In these cases, we give distance ranges rather
than uncertainties.

\section{RESULTS AND CONCLUSIONS}

Of the original 20 program stars, 4 were discarded due to strong stellar
features at the velocity of the cloud.  Seven stars 
were inconclusive and 
require higher SNR and/or higher spectral resolution than DDO
can provide in order to make use of them.  
Table \ref{table:program} lists the remaining 9 stars by
Tycho name, the apparent magnitude, spectral type, distance, equivalent
width, \ion{H}{1} column density due to the cloud, and note
regarding detection.

The DDO IVC Distance Project has been a resounding success and has far 
exceeded our initial expectations.  In particular the IVC 
$\mathrm{G}139.6+47.6$
had a relatively large number of background stars and few foreground
stars.  We determine the cloud's lower distance limit to be $129\pm10$ pc,
corresponding to the distance to $\mathrm{TYC}\ 4151\ 1452$.
Our closest detection star is $\mathrm{TYC}\ 4152\ 484$, which has spectral
class F5V, yielding an upper limit of $257^{+211}_{-30}$ pc.

Under the assumption that these distances are correct and given that every
star behind the cloud shows a detection, we conclude that the
covering factor is consistent with unity for this cloud.

To improve the distance determination, we would require more foreground
stars that are more distant, as well as background stars that are
closer than what is currently in our sample.  We have exhausted the
Tycho and USNO catalogs of appropriate stars.  It is possible
(though unlikely) that more candidates could be found by making a 
photometric survey of the region.  By considering stars bluer
than $B-V=0.35$, the survey would have to be complete to apparent magnitudes
of $m_{\rm V} = 10-11$ to sample the appropriate distances, assuming no 
extinction.
The Tycho Catalogue is
99\% complete down to $m_{\rm V} = 11$ \citep{Hog:1997} so it is unlikely 
that we have missed 
candidate stars with spectral type earlier than F.

The distance estimate might also be improved by further study of the
inconclusive stellar candidates.  In particular, better modeling of the
stellar features in the template spectra used for the cross-correlations
could reduce ambiguity and allow one to classify a detection or
non-detection with more confidence.

\acknowledgments{CRB, CT, MM, KB, MG, and AA wish to thank 
the Natural Sciences and Engineering
Research Council of Canada for support through postgraduate scholarships.
We wish to acknowledge the time and effort of observers 
directly related to the observations of this IVC:  
Wayne Barkhouse,
Mike Casey,
Tracy Clarke,
Heather Scott,
Quincy Kameda,
Jennifer Karr,
Patricia Mitchell,
Stefan Mochnacki,
Jason Rowe, and
Marcelo Ruetalo.  We also wish to thank the referee for their constructive
comments and prompt response.


\begin{figure}
   \plotone{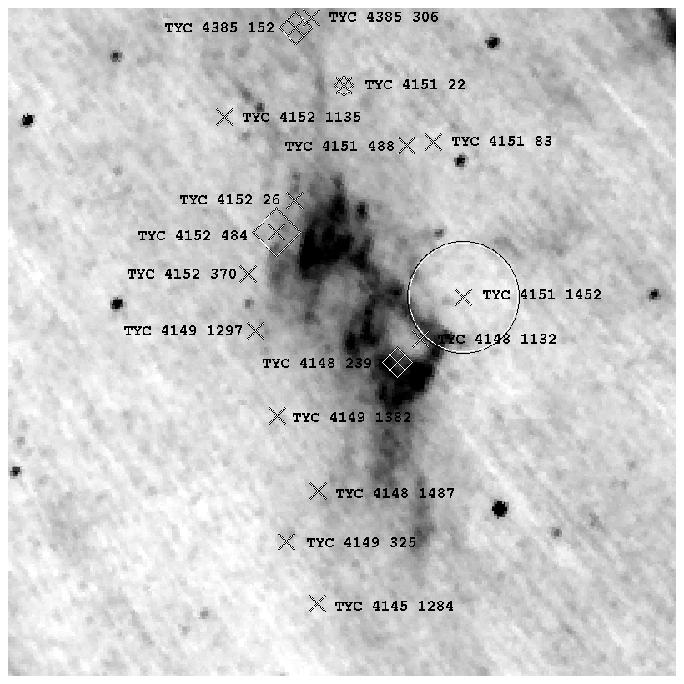}
   \caption{Positions of our target stars with respect to the IVC 
            cloud $\mathrm{G}139.6+47.6$.  A circle around a cross denotes a foreground
            star whereas a box around a cross denotes a background star.  Both
            boxes and circles are drawn with an angular size inversely 
            proportional to their nominal distances.  The background gray-scale
            is from the IRAS $100\mu$ survey.
            \label{fig:targets}}
\end{figure}

\begin{figure}
   \plotone{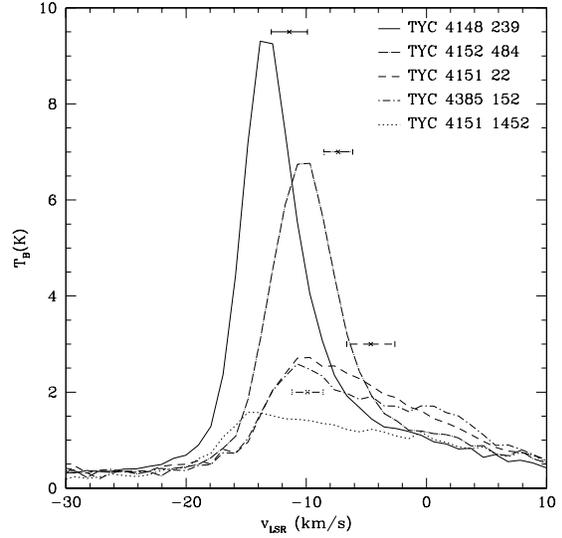}
   \caption{\ion{H}{1} emission along the line of sight to the four stars
            with detections and the one star with no detection 
            ($\mathrm{TYC}\ 4151\ 1452$).  The velocities
            of the detected absorption are plotted as error bars with line
            styles that match their stars' profiles.
            \label{fig:HI+vel}}
\end{figure}

\begin{figure}
   \plotone{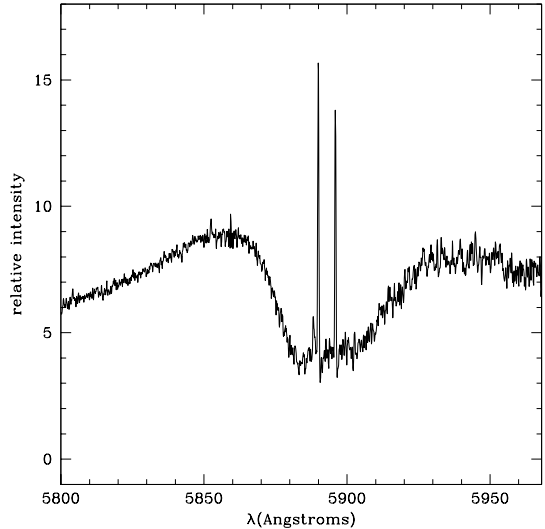}
   \caption{Intensity of the sky at the DDO, averaged over one month.  Note
            the self-absorption due to the high pressure sodium lights as well
            as the lines due to the low pressure lights.
            \label{fig:skylines}}
\end{figure}

\begin{figure}
   \plotone{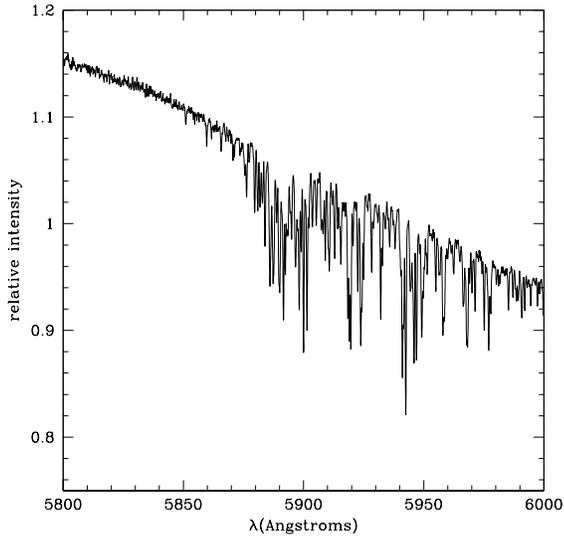}
   \caption{Sample spectrum of the telluric standard $\mathrm{HD}177724$.
            \label{fig:tell}}
\end{figure}

\begin{figure}
   \plotone{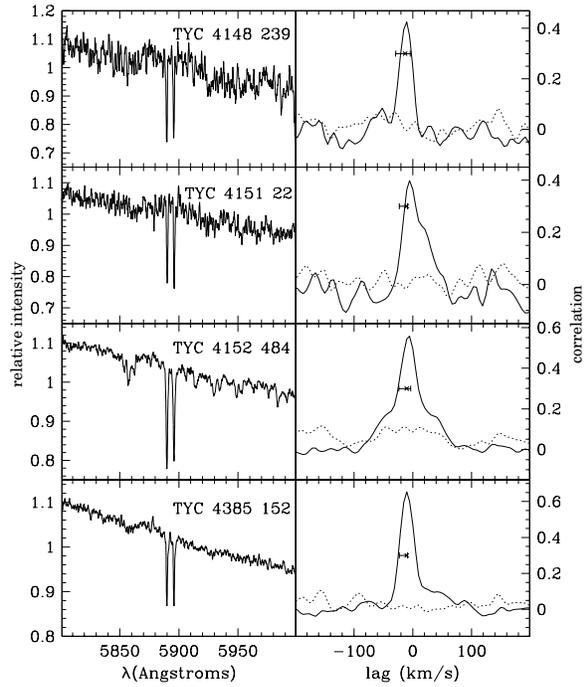}
   \caption{\ion{Na}{1} spectra and cross-correlations for the 
            four stars with detections.  Two cross-correlations are
            plotted for each star:  the sodium region, where only the
            \ion{Na}{1} lines are used (solid line) and the stellar region, 
            where
            all lines but the \ion{Na}{1} are used (dotted line). The cloud 
            velocities are 
	    over-plotted on the cross-correlations as error bars.
            \label{fig:cross_corr}}
\end{figure}

\begin{figure}
   \plotone{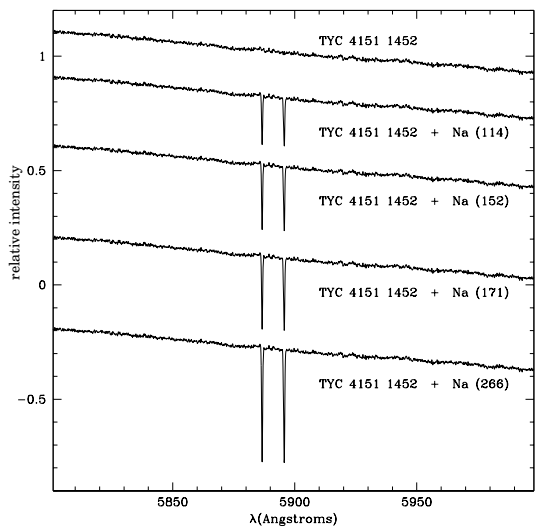}
   \caption{Expected \ion{Na}{1} absorption in the spectrum of
            $\mathrm{TYC}\ 4151\ 1452$ if it were behind the cloud.  The
            top-most spectrum is the observed spectrum.  Each subsequent
            spectrum is labeled with the EW(\ion{Na}{1})/$\rm N_{HI}$
            used to predict the equivalent width one would observe,
            which is shown in parentheses.
            \label{fig:fake_Na}}
\end{figure}

\begin{deluxetable}{lrrrrrc}
\tablecolumns{7}
\tablecaption{List of Program Stars. \label{table:program}}
\tablehead{\colhead{Tycho ID} & \colhead{$V$} & \colhead{MK Class}
           & \colhead{D} & \colhead{EW($\mathrm{D}_1$)} & 
           \colhead{$\mathrm{N}_{HI}$} & \colhead{Notes}\\
           \colhead{} & \colhead{(mag)} & \colhead{} & 
           \colhead{(parsecs)} & \colhead{EW($\mathrm{D}_2$)} & 
           \colhead{($\times 10^{19}\mathrm{cm}^{-2}$)}\\
           \colhead{} & \colhead{} & \colhead{} & 
           \colhead{} & \colhead{(m\AA)} & 
           \colhead{} & \colhead{}}
\startdata
TYC 4152 484 & 10.45&F5V & $257^{+211}_{-30}$ & $150 \pm 20$ & 10 & detection \\
 & & & & $130 \pm 20$ & & \\
TYC 4385 152 & 9.66 &A0V         & $649^{+206}_{-157}$ & $210 \pm 20$ & 26 & detection \\
 & & & & $200 \pm 20$ & & \\
TYC 4148 239 & 10.91&A1V         & $875_{-77}^{+449}$ & $310 \pm 40$ & 57 & detection \\
 & & & & $370 \pm 40$ & & \\
TYC 4151 22  & 10.53&A0V         & $968_{-233}^{+308}$ & $240 \pm 40$ & 28 & detection \\
 & & & & $270 \pm 40$ & & \\
TYC 4152 370 & 10.27&F0V         & $342_{-58}^{+176}$ & $380 \pm 50$ & 26 & possible detection \\
 & & & & $380 \pm 50$ & & \\
TYC 4145 1284& 10.45&F2(IV?)& $512^{+421}_{-207}$ & $330 \pm 70$ & 37 & possible detection \\
 & & & & $320 \pm 70$ & & \\
TYC 4149 325 & 9.8  &F2(?)      & $229 - 3000$ & $260 \pm 50$ & 14 & possible detection \\
 & & & & $250 \pm 50$ & & \\
TYC 4148 1132& 8.58 & $\ldots$     & $165 \pm 25$ & $\ldots$ & 47 & possible non-detection \\
 & & & & $\ldots$ & & \\
TYC 4151 1452& 6.40 & $\ldots$     & $129 \pm 10$ & $\ldots$ & 19 & non-detection \\
 & & & & $\ldots$ & & \\ \tableline
\enddata

\tablecomments{EW($\mathrm{D}_1$) and EW($\mathrm{D}_2$) are the
equivalent widths measured using the \ion{Na}{1} $\mathrm{D}_1$ and
$\mathrm{D}_2$ lines, respectively.}
\end{deluxetable}

\end{document}